Defects in Cd$_3$As$_2$ Epilayers Via Molecular Beam Epitaxy and Strategies for Reducing Them

A.D. Rice[1,*], K-W. Park[1,2], E.T. Hughes[3], K. Mukherjee[3], K. Alberi[1]

[1]National Renewable Energy Laboratory, Golden, CO 80401
[2]Division of Advanced Materials Engineering, Chonbuk National University,
Jeonju 54896, Republic of Korea
[3]Materials Department, University of California, Santa Barbara, California, 93106

**Abstract**

Molecular beam epitaxy offers an exciting avenue for investigating the behavior of topological semimetal Cd$_3$As$_2$, by providing routes for doping, alloying, strain engineering, and heterostructure formation. To date, however, minimal exploration has been devoted to the impact of defects that are incorporated into epilayers due to constraints imposed by the substrate and narrow growth window. Here, we use a combination of lattice-matched Zn$_x$Cd$_{1-x}$Te buffer layers, miscut substrates and broadband illumination to study how dislocations, twins and point defects influence the electron mobility of Cd$_3$As$_2$. A combination of defect suppression approaches produces Cd$_3$As$_2$ epilayers with electron mobilities upwards of 15,000 cm$^2$/V-s at room temperature.

## I. Introduction

Topological semimetals provide rich systems in which to explore new and interesting physics [1,2]. Their symmetry-protected band touching nodes with linear dispersions can host relativistic massless electrons at the Fermi level and support an array of transport behaviors not typically observed in trivial semiconductors or metals. Recent experimental investigations in Dirac and Weyl semimetals have demonstrated evidence for Fermi arcs [3], the quantum Hall effect [4], and giant magnetoresistance [5]. These breakthroughs will undoubtedly form the basis for future technologies, such as low-dissipation electronics and novel spintronics [6].

In parallel to these discovery efforts, it is critically important to develop the topological semimetal materials themselves. High quality crystals in bulk, epitaxial layer and nanostructure forms are necessary to enable future progress on both scientific and technological fronts. The Dirac semimetal Cd$_3$As$_2$ is a primary example [7]. Although many of the initial bandstructure and transport physics have been carried out in bulk single crystals [5,8], synthesis in epitaxial layers allows us to study the roles of confinement, strain, and heterostructures on their properties as well as fabricate them into devices.

Two factors strongly influence epitaxial growth of Cd$_3$As$_2$ and its resulting properties. The first is the substrate. Cd$_3$As$_2$ has a tetragonal crystal structure at temperatures < 475 °C with a I4$_1$/*acd* space group and lattice parameters a = 1.263 nm and c = 2.543 nm [9]. The atomic arrangement of its preferred (112) growth plane mirrors that of the (111) plane of zinc blende crystals [10]. Coincident atomic arrangements between the two can be achieved when the lattice constant of the zinc blende crystal is between those of ZnTe and CdTe. Previous growth on GaSb and CdTe substrates indicates that careful control of extended defects is necessary to achieve high electron mobilities [10,11]. Additionally, II-VI alloy substrates offer advantages over III-Vs due to their higher bandgaps, which limit them from introducing parallel

@Author to whom correspondence should be addressed:
Anthony.rice@nrel.gov

conduction channels. The second factor influencing epitaxy is the relatively narrow window of growth parameters that can be accessed. The high vapor pressure of $Cd_3As_2$ in vacuum practically restricts growth to temperatures below 250 °C, limiting the grower's ability to control surface morphology and defect formation. Even lower temperatures << 200 °C are necessary to maintain reasonable growth rates without loss to re-evaporation. In this context, approaches are needed to reduce point and extended defects generated as a result of heteroepitaxy or the use of low growth temperatures to realize high quality epitaxial structures that will enable future scientific and technological advances.

Here, we present a framework for $Cd_3As_2$ growth by molecular beam epitaxy (MBE) using elemental sources and lattice-engineered II-VI buffer structures. Relaxed $Zn_{1-x}Cd_xTe$ buffer layers on miscut GaAs(111) substrates spanning the entire solid solution provide a platform to control extended defects and strain conditions. Growth of $Cd_3As_2$ from elemental sources, allowing for flexibility in stoichiometry and defect control, is then demonstrated on these buffer structures. Finally, broadband light stimulation is explored as a route to further improve the electron mobility in $Cd_3As_2$ epilayers without raising the substrate temperature. The combination of these approaches ultimately improves the electron mobility and maximizes flexibility in designing the growth conditions of $Cd_3As_2$ epilayers, alloys and heterostructures for a variety of investigations.

## II. Experimental Methods

Epilayers were grown in an interconnected, dual chamber Omicron EVO25 MBE system. All group II, III and VI sources were evaporated from effusion cells, while $As_4$ was evaporated from cracker sources in both chambers. Substrate temperatures were calibrated using a k-space band-edge thermometry system. Oxide removal of GaAs(111) under $As_4$ was performed in a dedicated III-V chamber followed by growth of a 300-nm GaAs buffer layer at high temperatures (>610°C) and low V/III ratios (3-5:1) [12]. The substrates were then cooled down and capped with an arsenic film. Samples were subsequently removed from vacuum and re-loaded on Mo blocks specific to the II-VI chamber to minimize contamination. The arsenic cap was thermally desorbed at ~350 °C, and the substrate was cooled down to commence II-VI and II-V growths. Specific growth conditions are described below. Unless noted, the sample surface was illuminated with broadband light from a Xe lamp (Oriel 6255) during growth of both the II-V buffer and $Cd_3As_2$ epilayers. The lamp powers are stated below. Hall measurements were performed at room temperature using annealed indium contacts and Van der Pauw geometry under excitation voltages of 5 mV (50-200 μA). Multiple samples were measured for each growth condition to produce the mobility ranges reported in the results section. Electron-channeling contrast imaging (ECCI) was used to probe for structural defects in the $Cd_3As_2$ epilaylers [13]. Images were taken in the $Cd_3As_2$ (-408)/(-440) channeling condition to characterize threading dislocation defect populations and other extended defects.

### III. **Results**

We first discuss the properties of the lattice-engineered II-VI buffer layers, as they will affect the $Cd_3As_2$ morphology and electron mobility. Figure 1 shows reflective high energy electron diffraction (RHEED) patterns and atomic force microscopy (AFM) scans from typical buffer structures. All buffers started with a ZnTe epilayer (7.3% mismatch with GaAs), grown directly on on-axis GaAs(111) using a Zn treatment prior to growth to maximize the ZnTe crystalline quality [14]. Growth using a VI/II ratio of 2:1, a substrate temperature of 330 °C and broadband illumination at a lamp power of 100 W produced the smoothest films and highest intensity thickness fringes in x-ray diffraction (XRD) (Fig. 1a), consistent with sharp interfaces. Stoichiometric or group II rich growths resulted in very spotty RHEED patterns (not shown). Growing these epilayers thicker than 25 nm was also found to result in rougher morphologies. After ZnTe, a final $Zn_xCd_{1-x}Te$ epilayer was grown. For the extreme endpoint of CdTe, an approximately 150 nm layer was grown on the relaxed ZnTe surface under similar flux conditions at a lower substrate temperature (~240 °C), resulting in film roughnesses comparable to the starting GaAs layers. $Zn_xCd_{1-x}Te$ epilayers of mixed composition were grown at intermediate temperatures depending upon the relative Zn and Cd content. For instance, a Cd/Zn beam flux ratio of ~2:1 and a substrate temperature of 290 °C were used to achieve a lattice matching composition ($Zn_{0.42}Cd_{0.58}Te$) to the $Cd_3As_2$ epilayer.

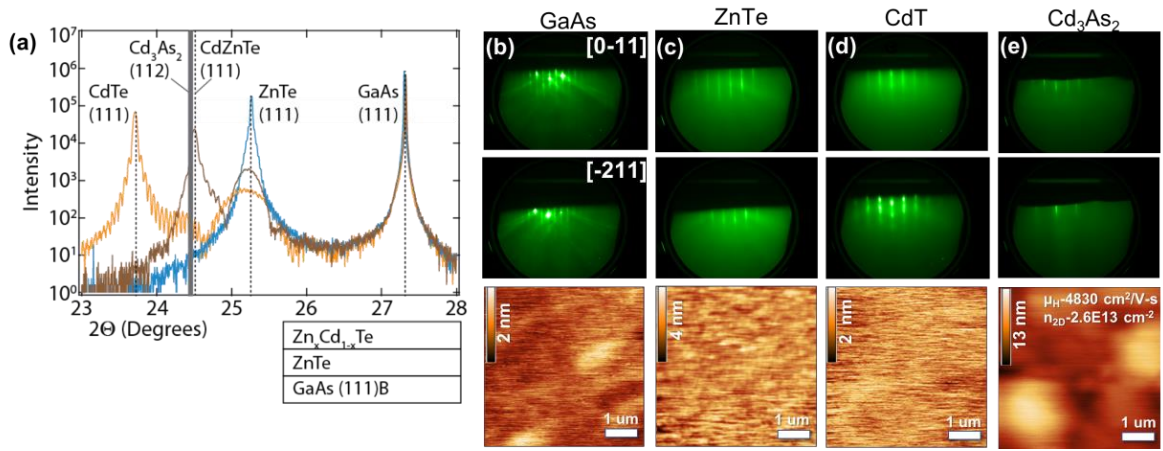

**Figure 1**.a) XRD and b) AFM and RHEED images of a $Cd_3As_2$/ZnCdTe/ZnTe/GaAs buffer structures. All GaAs substrate orientations were on-axis.

We next discuss the properties of the $Cd_3As_2$ epilayers grown on these II-VI buffers. Table I summarizes the impact of the buffer composition, substrate miscut and broadband illumination on the $Cd_3As_2$ electron mobility. $Cd_3As_2$ growth was performed at substrate temperatures of approximately 100-120 °C under $As_4$ rich conditions. These low temperatures are required due to the high vapor pressure of $Cd_3As_2$ [15]. Slightly higher substrate temperatures were previously found to produce higher electron mobilities, where a nearly two-fold increase was observed when raising the temperature from 140 °C to 210 °C [10]. However, these temperatures also require high flux rates to overcome elevated $Cd_3As_2$ decomposition, so we therefore utilized much lower temperatures that are more practical for growth. All growths under illumination conditions were carried out with a Xe lamp setting of 100 W except in the power-dependence study discussed later.

**Table I.** Summary of $Cd_3As_2$ electron mobility based on substrate, buffer, and growth conditions

| Growth Condition | | | Electron Mobility [cm2/V-s] |
|---|---|---|---|
| **Substrate Miscut** | **II-VI Buffer** | **Illumination** | |
| On axis | CdTe | Dark | 2,000-4000 |
| On axis | CdTe | Illuminated | 4,000-5,000 |
| 3° miscut | CdTe or ZnTe | Illumianted | 7,000-9,000 |
| 3° miscut | ZnCdTe | Dark | 7,000-8,000 |
| 3° miscut | ZnCdTe | Illuminated | 13,000-15,000 |

Figure 1e shows RHEED and AFM results from Cd$_3$As$_2$ epilayers grown on a lattice-mismatched CdTe-terminated buffer on an on-axis GaAs substrate. CdTe layers showed smoother morphologies and narrower diffraction peaks than ZnTe and also allows for comparison with previous reports [10,16]. A 4x reconstruction is observed parallel to the [0-11] and [-211] directions (relative to GaAs(111)). A smooth transition between the II-VI and Cd$_3$As$_2$ is observed, with no evidence of roughening during film nucleation. Previous reports have not discussed the surface reconstruction explicitly, but published images appear to show faint 4x streaks [10,16] The measured lattice parameter in XRD is in good agreement with reported values, while AFM revealed 2D but rough morphologies relative to the underlying layers (~2-3 nm r.m.s. roughness). Epilayer thicknesses were not directly measured, but x-ray reflectivity performed on thinner samples suggest the Cd$_3$As$_2$ epilayers presented here were approximately 200-250 nm thick. Baseline Cd$_3$As$_2$ epilayers grown under dark conditions on on-axis substrates with lattice-mismatched CdTe buffers exhibit low electron mobilities ~2,000-4,000 cm$^2$/V-s. The electron mobilities improved to 4,000-5,000 cm$^2$/V-s when broadband illumination was applied. All films, regardless of the illumination conditions, had sheet carrier concentrations of 3-6E13 cm$^{-2}$, which corresponds to 1-3E18 cm$^{-3}$ given approximate the film thicknesses.

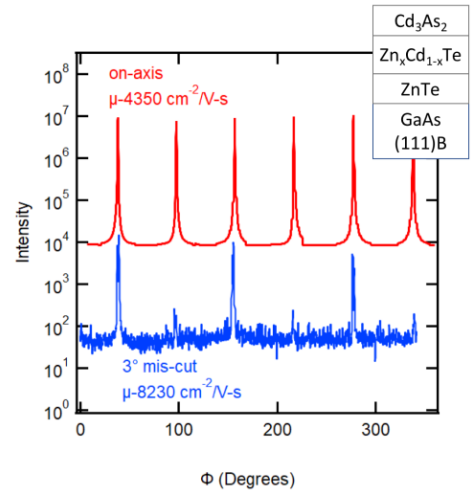

**Figure 2.** Phi scans of Cd3As2 (44$\underline{16}$) peaks grown on varying miscut substrates and ZnCdTe buffers.

Previous reports showed mild twinning arising at the Cd$_3$As$_2$ interface [10,11]. Twinning is also a known issue in CdTe epilayers [17]. Figure 2 displays phi scans of the (44$\underline{16}$) peaks of Cd$_3$As$_2$ grown on Zn$_x$Cd$_{1-x}$Te buffers. The six-fold symmetry of these peaks arises from the 180° rotations that occur during twinning. Similar degrees of twinning are observed in both Cd$_3$As$_2$ layers (red curve) and Zn$_x$Cd$_{1-x}$Te layers across the entire solid solution grown on on-axis substrates (not shown), suggesting twinning is largely initiated in the buffer layers and then propagates throughout the structure. Twinning was substantially reduced by performing growths under similar conditions on wafers 3° mis-cut toward (2-1-1), as shown in Fig. 2, similar to previous reports [11]. In particular, little evidence of twins was observed in both the Cd$_3$As$_2$ (blue curve) or Zn$_{0.42}$Cd$_{0.58}$Te layers (not shown) with nearly lattice-matched compositions.

Figure 3 summarizes the impact of the II-VI buffer lattice constant on the Cd$_3$As$_2$ properties. Reciprocal space maps confirm that the Cd$_3$As$_2$ epilayers are highly relaxed on 3° miscut CdTe (2.3% mismatch) and ZnTe (-3.4% mismatch) buffers, while Cd$_3$As$_2$ grown on Zn$_{0.42}$Cd$_{0.58}$Te buffers appeared nearly lattice matched with narrower peak shapes. No significant mobility difference is observed between epilayers grown on either ZnTe and CdTe buffers, despite the lower Cd$_3$As$_2$ surface roughness on CdTe. This suggests that morphology only plays a minor role in limiting the electron mobility compared to other defects, which is consistent with previous reports [10]. Cd$_3$As$_2$ epilayers grown on lattice-matched Zn$_{0.42}$Cd$_{0.58}$Te buffers exhibited similarly rough surfaces to relaxed Cd$_3$As$_2$ epilayers grown on CdTe buffers. To evaluate the influence of the II-VI buffer lattice constant on extended defects in Cd$_3$As$_2$, we performed ECCI measurements on Cd$_3$As$_2$ epilayers grown on all three buffers. The micrographs (Fig. 3) indicate that the epilayers have similar threading dislocation densities (~4-6 x 10$^8$ cm$^{-2}$); however, there is a substantial reduction in the density of other extended defects in the epilayers grown on lattice-matched Zn$_x$Cd$_{1-x}$Te buffers compared to those grown on lattice-mismatched ZnTe and CdTe buffers. The former is more structurally homogeneous but does show signs of pitting at the film surface. Although we are still exploring the origin of these differences, reducing the lattice mismatch between the buffer and epilayer appears to help suppress the formation of certain types of extended defects.

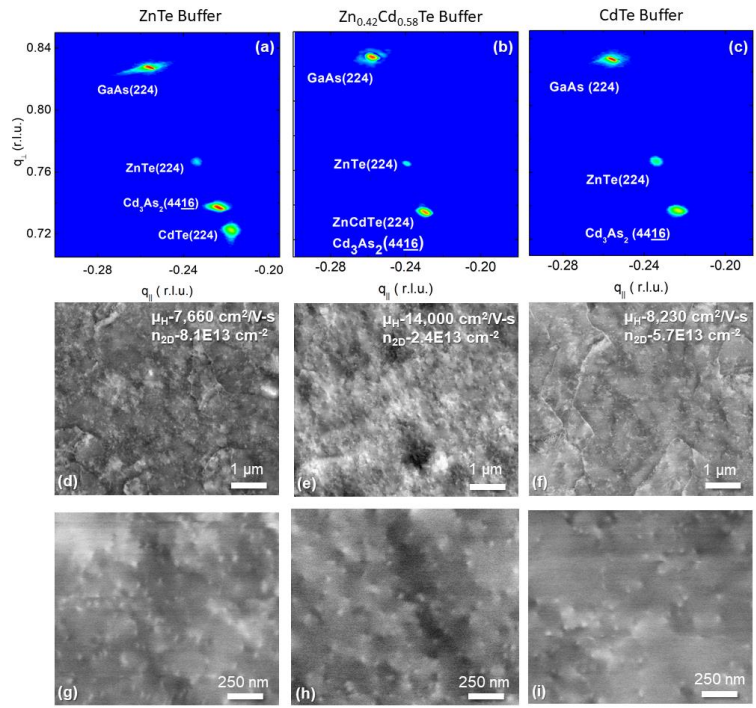

**Figure 3**. (top) Reciprocal space maps and (middle and bottom) ECCI images of Cd$_3$As$_2$ grown on ZnTe, ZnCdTe, and CdTe buffer structures. Electron mobility and sheet density density from room temperature Hall measurements are noted in the middle panels.

Suppression of extended defects has a substantial effect on the electron mobility. When twinning is reduced through the use of a miscut substrate but a lattice-mismatched CdTe or ZnTe buffer is used, the electron mobility improves to 7,000-9000 cm$^2$/V-s. The addition of a lattice-matched Zn$_{0.42}$Cd$_{0.58}$Te buffer further increases the mobility to 13,000-15,000 cm$^2$/V-s. These values are close to the highest reported room temperature values for epitaxially grown thin films (19,300 cm$^2$/V-s [10]).

Given the previous success broadband illumination of the growth surface has had on improving the optical properties of II-VI semiconductors grown at low temperature [18], we explore it as a tool for further improving epitaxial Cd$_3$As$_2$. All epilayers discussed up to this point were grown under broadband illumination with 100 W power. Table II details the impact of Xe lamp power on the electron mobility and surface roughness of Cd$_3$As$_2$ epilayers grown on lattice-matched, miscut Zn$_{0.42}$Cd$_{0.58}$Te buffers. While the film morphology became rougher with increasing illumination, the electron mobility increased up to intermediate lamp powers of 100 W. Illumination of semiconductor surfaces during growth has been demonstrated to reduce point defects, which is likely a particular concern here due to the low substrate temperatures used for Cd$_3$As$_2$ growth. Light has previously been found to affect semiconductor growth by increasing adatom desorption [19], altering surface kinetics [20] and modifying defect formation [21]. Changes in surface kinetics under illumination could also be responsible for the surface roughening [22]. Both Cd and As have non-unity sticking coefficients, so it is possible the net result of illumination on this surface is more Cd and/or As desorption. Reduction in defects by light stimulation would cause an increase in the electron mobility, although the exact pathways and the specific defect population affected have not been identified at this time. Further investigation is required to understand the underlying mechanisms of both trends in this material system.

Table II. Summary of the effect of Xe lamp power on the electron mobility and sheet density and the surface roughness of lattice-matched Cd$_3$As$_2$ epilayers. All sample had a sheet carrier concentration of 4-5E13 cm$^{-2}$

| Lamp power [W] | Electron Mobility [cm2/V-s] | r.m.s Roughness [nm] |
|---|---|---|
| 0 | 7670 | 1.2 |
| 20 | 8280 | 1.3 |
| 100 | 13500 | 1.5 |
| 150 | 10600 | 2.1 |

While improvements in Cd$_3$As$_2$ electron mobility were gained through the use of lattice-matched II-VI buffers, miscut substrates and broadband illumination, only a combination of the three produced electron mobilities above 10,000 cm$^2$/V-s. Consistent with Matthiessen's rule, our results illustrate that reducing the density of one type of defect can only improve the electron mobility of Cd$_3$As$_2$ so much when relatively large densities of others exist. This rule is evident in the magnitude of the improvement produced by broadband light illumination. When growth is carried out on on-axis substrates and lattice-mismatched buffers, illumination only increased the electron mobility by ~1,000 cm$^2$/V-s, suggesting the mobility is dominated by extended defects. However, when illumination was applied to growth on miscut substrates and lattice-matched Zn$_{0.42}$Cd$_{0.58}$Te buffers, illumination produced a 5,000 cm$^2$/V-s increase. This result suggests that illumination affects some additional type of defect population.

Overall, our study suggests that multiple types of defects are the main cause of the low electron mobilities in epitaxially-grown Cd$_3$As$_2$ compared to high quality bulk single crystals (42,850 cm$^2$/V-s [23]). We suggest the following strategies to realize further gains. The dislocation densities in our Cd$_3$As$_2$ epilayers (~10$^8$ cm$^{-2}$) are still quite high and will need to be reduced through better design of the II-VI buffers. Disorder and point defects can be reduced through annealing above the practical growth temperatures used here with a sufficiently stable capping layer [24]. The role of the II/V ratio must also be explored to obtain additional insight into the role of Cd or As vacancies (and other native defects) on the electron mobilities. However, the higher symmetry of the zinc blende structure will ultimately be a limiting factor at some point.

IV.     Conclusions

In summary, we demonstrate a II-VI platform providing tuneable lattice constants and smooth morphologies for $Cd_3As_2$ growth. Controlling defects using miscut substrates, lattice-matched buffer layers and broadband illumination produced higher electron mobilities while maintaining lower growth temperatures to limit $Cd_3As_2$ decomposition. Our results provide insight into the types of defects formed within epitaxial $Cd_3As_2$ and their influence on the transport properties. Such studies are needed as we push the limits of epitaxial crystal quality to enable new physics discoveries within the $Cd_3As_2$ material system and utilize it for future techological advances.

**Acknowledgements**

This work was authored in part by Alliance for Sustainable Energy, LLC, the Manager and Operator of the National Renewable Energy Laboratory for the U.S. Department of Energy (DOE) under Contract No. DE-AC36-08GO28308. Funding was provided by the Laboratory Directed Research and Development program. The views expressed in the article do not necessarily represent the views of the DOE or the U.S. Government. The U.S. Government retains and the publisher, by accepting the article for publication, acknowledges that the U.S. Government retains a nonexclusive, paid-up, irrevocable, worldwide license to publish or reproduce the published form of this work, or allow others to do so, for U.S. Government purposes.